\documentclass[cameraready]{Interspeech}

\title{VieSpeaker: A Large-Scale Vietnamese Speaker Recognition Dataset Beyond Visual Dependency}

\author[equalcontribution]{Viet Hoang}{Pham}
\author[equalcontribution]{Tran Trung}{Nguyen}
\author[equalcontribution]{Bao Thu}{Ho}

\author[]{Phuong Tuan}{Dat}
\author[correspondingauthor]{Thi Thu Trang}{Nguyen}

\address{
    Hanoi University of Science and Technology, Hanoi, Vietnam
}

\email{
\{hoang.pv224854, trung.nt2400117, thu.hb226003\}@sis.hust.edu.vn, \\
phuongtuandat2915@gmail.com, trangntt@soict.hust.edu.vn
}


\usepackage{comment}
\usepackage{booktabs}
\usepackage{diagbox}
\usepackage{adjustbox}
\usepackage{mdframed}

\usepackage{multirow}

\begin{document}

\maketitle

\begin{abstract}

Speaker recognition has advanced rapidly with large-scale training datasets, yet Vietnamese remains under-resourced, with existing corpora limited in scale and acoustic diversity. Most large-scale datasets rely on facial cues to link speech with speaker identities, restricting data collection to recordings where speakers appear on camera. We propose a face-independent dataset construction pipeline and introduce VieSpeaker, a large-scale Vietnamese speaker recognition dataset. Our approach leverages textual metadata and large language model reasoning to infer speaker identities from transcripts and contextual information. VieSpeaker contains approximately 902 hours of speech from 4,715 speakers. Experiments show that models trained on VieSpeaker achieve improved robustness and generalization compared to existing Vietnamese datasets. This work demonstrates the feasibility of face-independent dataset construction and provides a new direction for building large-scale speech resources.
\end{abstract}


\section{Introduction}
Speaker recognition refers to the automatic analysis of speech signals to determine a speaker’s identity or verify a claimed identity. Recently, significant progress in speaker recognition has been largely driven by the availability of large-scale training corpora, which enable deep neural networks to effectively model both intra-speaker variability and inter-speaker diversity. Datasets such as VoxCeleb2~\cite{vox2} and CN-Celeb2~\cite{cnceleb2} contain millions of utterances from thousands of speakers recorded in unconstrained real-world environments, providing high speaker diversity, rich acoustic variability, and sufficient data volume to learn robust and discriminative speaker embeddings.

Compared to well-resourced languages such as English or Chinese, Vietnamese speaker recognition remains relatively data-scarce, with only small- to medium-scale datasets available \cite{vlsp, vietnameceleb, Voxvn, vsasv}.
Among them, Vietnam-Celeb \cite{vietnameceleb} is one of the earliest benchmarks for Vietnamese speaker recognition. However, it mainly contains recordings of public figures collected from interviews or studio environments, resulting in a relatively limited speaker coverage and recording diversity. This constraint can lead to significant performance degradation when models are evaluated under cross-domain or mismatched conditions, as reported in \cite{ood_degrade_1, ood_degrade_2, ood_degrade_3}.

More recently, VoxVietnam \cite{Voxvn} alleviates this issue by removing the manual speaker selection and adopting deep clustering approach for speaker discovery. Nevertheless, its construction pipeline still relies on face detection and face tracking to determine speaker identity, a limitation shared by many existing speaker datasets \cite{vox2, cnceleb2}. 
Although effective for scalable labeling, such visual supervision imposes a strong multimodal constraint, excluding recordings where speakers are not visible (e.g., vlogs, anonymous podcasts, or phone conversations) even when the audio remains identity-consistent. Moreover, reliance on facial cues introduces additional data acquisition costs due to image collection and makes annotation quality sensitive to image conditions such as pose, lighting, and resolution. This limitation is less pronounced in large-scale datasets due to their massive scale, whereas VoxVietnam remains constrained by comparatively limited data resources.

To address these limitations, we propose a novel pipeline for constructing large-scale Vietnamese speaker datasets that eliminates the need for visual modality in identity supervision. Instead, our approach leverages recent advances in Large Language Models (LLMs) to perform structured reasoning over textual metadata (titles, channel names, and descriptions), inspired by prior studies \cite{llmreason_1, llmreason_2, llmreason_3}. This modification enables the inclusion of speech that was previously excluded while expanding the speaker pool beyond public figures. Rather than replacing face-dependent pipelines, the proposed approach offers an alternative pathway for speaker identity inference and opens the possibility of future hybrid approaches. As a result, we establish the first large-scale Vietnamese speaker recognition dataset with substantially greater speaker diversity and richer acoustic variability, providing a valuable resource for developing and evaluating Vietnamese speaker recognition systems.

The remainder of this paper is organized as follows. 
Section~\ref{sec:pipeline} introduces
the data construction pipeline.
Section~\ref{sec:dataset} presents the resulting dataset statistics. 
Section~\ref{sec:experiments} reports experimental evaluations on speaker recognition tasks. 
Finally, Section~\ref{sec:conclusion} concludes the paper and discusses future directions.

\begin{figure*}[t]
    \centering
    \includegraphics[width=1.15\textwidth]{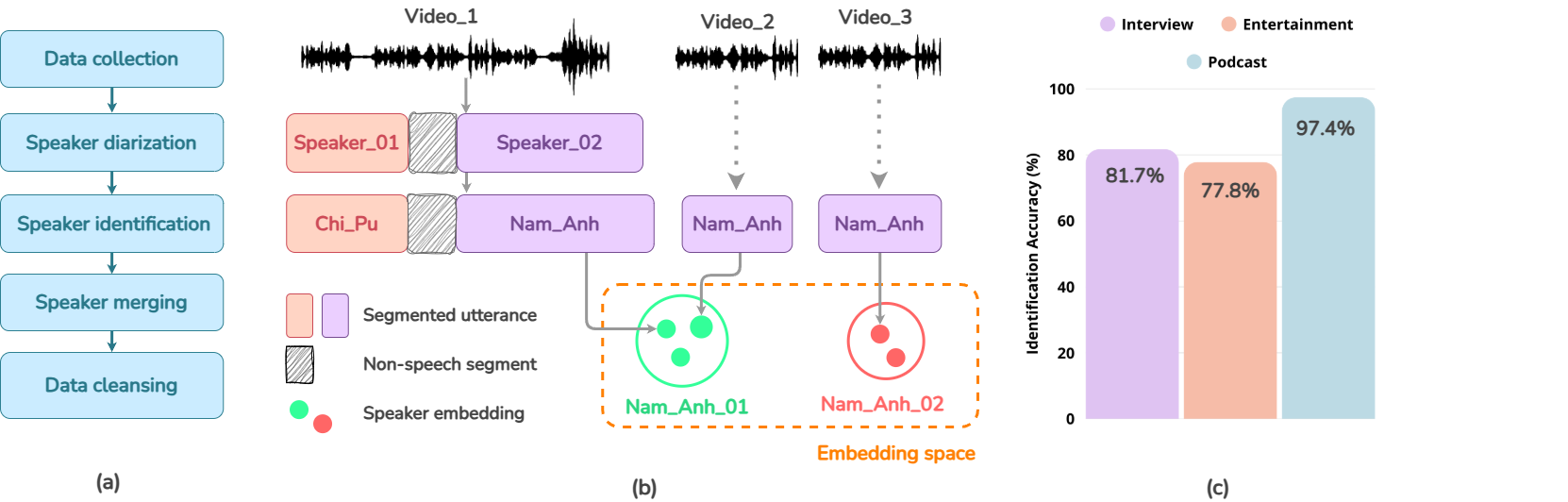}
    \caption{Overview of the proposed dataset construction pipeline. (a) Main processing workflow. (b)  Illustration of stage-wise outputs. (c) Speaker identity coverage from metadata}
    \label{fig:pipeline}
\end{figure*}
\section{Dataset construction pipeline}
\label{sec:pipeline}
In this section, we describe the proposed dataset construction pipeline, as illustrated in Fig.~\ref{fig:pipeline}, and detail each of its key stages.
\subsection{Data collection}

Our data collection begins with manually curating playlists from publicly accessible YouTube channels across three domains: interviews, entertainment, and podcasts. Unlike prior pipelines that are largely restricted to on-camera studio recordings, our face-independent design enables broader coverage within these domains, including radio interviews, call-in programs, audio-centric podcasts, and diverse game show formats where visual cues are absent or inconsistent. An automated processing pipeline retrieves publicly available video links and associated metadata, including descriptions and transcripts, which are used for downstream speaker identity reasoning and validation. Raw media content remains hosted on the original platform and is not redistributed as part of the released dataset.

\subsection{Speaker diarization}
We adopt a speaker diarization framework to segment raw audio recordings into speaker-homogeneous units. Specifically, the \texttt{speaker-diarization-3.1} pretrained checkpoint from the Pyannote framework\footnote{\url{https://huggingface.co/pyannote/speaker-diarization-3.1}} is employed to generate time-stamped utterance boundaries and assign anonymous speaker identifiers, denoted as \texttt{SPEAKER\_ID}. These local identifiers then support identity resolution using metadata and lexical evidence.

\subsection{Speaker identification}
\label{sec:reasoning}
Following diarization, we introduce an LLM-driven module implemented with the \texttt{gemini-2.5-pro} model via the Google AI Studio API. The model performs reasoning over the transcripts of each \texttt{SPEAKER\_ID} together with the corresponding video metadata (title, channel name, and description).

To ensure stable and evidence-driven identity inference, we design a structured prompt with controlled decoding. Each prompt processes all diarized speaker groups from a single video, where up to seven initial transcript segments are sampled for each \texttt{SPEAKER\_ID}. Minimal preprocessing is applied, and decoding is performed deterministically using temperature=0.0 and top\_k=1. A simplified prompt template is shown below:

\begin{mdframed}
\scriptsize
You are an expert audio analyst identifying speakers in Vietnamese multimedia recordings. You are given:

VIDEO METADATA: title, channel name, and description.

TRANSCRIPT SAMPLES: diarized segments grouped by SPEAKER\_ID, formatted as:

[start\_time] [end\_time] [SPEAKER\_ID]

\textbf{Task:} Map each \texttt{SPEAKER\_ID} to a real-world identity.  

\textbf{Constraints:}  
(1) Assign a name only if explicit textual evidence is present (e.g., self-introduction or direct addressing).  
(2) Do not guess; output ``Unknown'' when uncertain.  
(3) Merge \texttt{SPEAKER\_ID}s if strong evidence indicates they belong to the same individual.  
(4) Output verified Vietnamese names with correct diacritics.  

\textbf{Output:} A single JSON object mapping each \texttt{SPEAKER\_ID} to \{name, evidence\}.
\end{mdframed}

This design serves two purposes. First, it enforces an evidence-based annotation strategy, minimizing false identity assignments. Second, by allowing identity consolidation across semantically consistent segments, the module implicitly refines diarization outputs by correcting over-segmentation errors. The final output of this stage is a verified mapping from each \texttt{SPEAKER\_ID} to its resolved identity, accompanied by extractive textual justification.
\subsection{Speaker merging}

In practice, a single speaker, particularly celebrities, can appear across multiple playlists, making identity consolidation necessary to avoid fragmented labels. We first normalize the detected speaker names by removing role-based prefixes (e.g., “ca sĩ Chi Pu” to “Chi Pu”) and standardizing formatting variations (e.g., “Chi-Pu”, “CHI PU”) through an additional LLM-based step. This process ensures that different surface forms are mapped to a single canonical identity. Finally, we manually remove foreign speakers to retain only target-language identities.

The remaining speakers are grouped by their canonical names. For identities spanning multiple videos, we perform cosine-similarity-based agglomerative clustering using ECAPA-TDNN \cite{ecapa} embeddings trained on Vietnam-Celeb. Derived from the similarity distribution of a small clean validation subset, the \textit{merge} and \textit{split} thresholds are set to 0.7 and 0.2, respectively. Consequently, highly similar pairs are merged, dissimilar ones separated, and ambiguous cases (including aliases or nicknames) are human-verified to ensure consistency. Finally, to evaluate metadata-driven identity recovery, we measure the proportion of identities successfully recovered from metadata across the three genres (Fig.~\ref{fig:pipeline}(c)).

\subsection{Data cleansing}
Following \cite{outlier}, we remove anomalous utterances using the Interquartile Range (IQR) on cosine similarity scores to the global mean embedding. Finally, utterances shorter than 1.0 second are discarded, and speakers with a total speech duration below 30 seconds are excluded to ensure sufficient data per speaker.
For dominant speakers such as hosts, we randomly downsample their utterances when their cumulative duration becomes excessively large. This mitigates bias toward overrepresented speakers and promotes a more balanced training distribution.
\section{Data description}
\label{sec:dataset}
After completing the proposed data construction process, we obtain the finalized VieSpeaker dataset comprising 365,874 utterances from 4,715 unique speakers, totaling 902.03 hours of speech. VieSpeaker becomes the largest and most comprehensive dataset for Vietnamese speaker recognition to date. 

\subsection{Utterance and genre distribution}
\begin{table}[t]
  \caption{Utterance duration distribution of VieSpeaker.}
  \label{tab:utterance_dist}
  \centering
  \begin{tabular}{|lrr|}
    \hline
    \textbf{Duration (s)} & \textbf{\# of Utterances} & \textbf{Proportion (\%)} \\
    \hline
    \hline
    $<$ 2   & 34,610 & 9.46 \\
    2 -- 5  & 97,785 & 26.73 \\
    5 -- 10 & 103,779 & 28.36 \\
    10 -- 20 & 99,186 & 27.11 \\
    $>$ 20  & 30,514 & 8.34 \\
    \hline
    \hline
    \textbf{Total} & \textbf{365,874} & \textbf{100.00} \\
    \hline
  \end{tabular}
\end{table}
\begin{table}[b]
  \caption{The genre distribution of VieSpeaker.}
  \label{tab:genre_dist}
  \centering
  \begin{tabular}{|lrrr|}
    \hline
    \textbf{Genre} & \textbf{\# of Spks} & \textbf{\# of Utterances} & \textbf{\# of Hours} \\
    \hline \hline
    Interview & 927 & 236,461 & 623.23 \\
    Entertainment & 3,723 & 107,368 & 217.41 \\
    Podcast & 94 & 22,045 & 61.38 \\
    \hline \hline
    \textbf{Total} & \textbf{4,715} & \textbf{365,874} & \textbf{902.03} \\
    \hline
  \end{tabular}
\end{table}

Table \ref{tab:utterance_dist} summarizes the utterance length distribution. Unlike earlier corpora dominated by very short segments, VieSpeaker shows a more balanced profile, with over half of the data between 5 and 20 seconds. This range provides richer phonetic and acoustic variability for training, especially when random chunk sampling is applied.

Table \ref{tab:genre_dist} shows the distribution of speakers, utterances, and total duration across three domains. The entertainment domain has the most speakers but less total speech, reflecting interactive formats. In contrast, the interview domain contains fewer speakers yet contributes most of the speech hours, while podcasts provide long, uninterrupted segments with consistent recording conditions.

\subsection{Comparison with other datasets}
Table \ref{tab:comparison} compares VieSpeaker with existing Vietnamese and international speaker recognition corpora. 
Compared to Vietnam-Celeb and VoxVietnam, VieSpeaker represents a substantial leap in scale, expanding the speaker pool more than three times and increasing the total duration by nearly five-fold relative to earlier Vietnamese resources.  When compared with established speaker recognition benchmarks, VieSpeaker meaningfully narrows the resource gap. In particular, compared with CN-Celeb2, it contains more than twice the number of speakers while achieving a comparable total duration, placing Vietnamese resources at a more competitive level.
Although still smaller than VoxCeleb2, which benefits from far larger global data availability, VieSpeaker establishes a solid benchmark for unconstrained Vietnamese speaker recognition.
Overall, VieSpeaker advances the scale and maturity of Vietnamese speaker recognition datasets and provides a strong basis for future expansion in both duration and genre coverage.

\begin{table}[t]
  \caption{Comparison with other large-scale speaker datasets.}
  \label{tab:comparison}
  \centering
  \begin{tabular}{|lrrr|}
    \hline
    \textbf{Name} & \textbf{\# of Spks} & \textbf{\# of Utterances} & \textbf{\# of Hours} \\
    \hline
    \hline
    VoxCeleb1~\cite{vox1} & 1,251 & 153,516 & 352 \\
    VoxCeleb2~\cite{vox2} & 6,112 & 1,128,246 & 2,794 \\
    CN-Celeb1~\cite{cnceleb1}& 1,000 & 130,109 & 274 \\
    CN-Celeb2~\cite{cnceleb2}& 2,000 & 529,485 & 1,090 \\
    \hline \hline
    Vietnam-Celeb~\cite{vietnameceleb}& 1,000 & 87,140 & 187 \\
    VoxVietnam~\cite{Voxvn}& 1,406 & 187,980 & 261 \\
    \textbf{VieSpeaker} & \textbf{4,715} & \textbf{365,874} & \textbf{902} \\
    \hline
  \end{tabular}
\end{table}
\begin{table}[b]
\centering
\caption{Statistics of the final dataset subsets.}
\label{tab:subset}
\begin{tabular}{|lrrr|}
\hline
\textbf{Subset} & \textbf{\# of Spks} & \textbf{\# of Utterances} & \textbf{\# of Pairs} \\
\hline \hline
VieSpeaker-T & 4,000 & 320,527 & -- \\
VieSpeaker-E & 715 & 45,319 & 500,000 \\
VieSpeaker-H & 715 & 45,253 & 500,000 \\
\hline
\end{tabular}
\end{table}
\subsection{Final Dataset}
\begin{table*}[!t]
\centering
\caption{EER (\%) on Vietnam-Celeb and VoxVietnam benchmarks under different training and pretraining settings.}
\label{tab:cross_eval}
\renewcommand{\arraystretch}{1.2}
\begin{tabular}{|l|c|c|c|c|}
\hline
\diagbox[width=15em]{\textbf{Training set}}{\textbf{Test set}} 
& \textbf{Vietnam-Celeb-E }
& \textbf{Vietnam-Celeb-H }
& \textbf{VoxVietnam-E}
& \textbf{VoxVietnam-H}\\
\hline\hline
VoxCeleb2 & 14.79 & 17.80 & 21.00&	28.06 \\
Vietnam-Celeb-T & 6.53 & 7.92 & 13.48&	22.59 \\
VoxVietnam-T & 15.05 & 16.67 & 13.66&	22.26\\
VieSpeaker-T & 9.28 & 11.19 & 13.19&	21.78 \\
VoxCeleb2 ft. Vietnam-Celeb-T & \underline{5.79} & \underline{6.91} & 13.36&	22.06 \\
VieSpeaker-T ft. Vietnam-Celeb-T & \textbf{5.45} & \textbf{6.74} & 13.28&	22.00 \\
VoxCeleb2 ft. VoxVietnam-T & 11.02	&11.99&	\underline{12.70}&	\textbf{21.41} \\
VieSpeaker-T ft. VoxVietnam-T&9.97	&11.08&	\textbf{12.65}&	\underline{21.43}\\
\hline
\end{tabular}
\end{table*}

\begin{table}[t]
\centering
\caption{EER (\%) on VieSpeaker test protocols.}
\label{tab:vs_eval}
\renewcommand{\arraystretch}{1.2}
\setlength{\tabcolsep}{8pt}
\begin{tabular}{|p{4.6cm}|cc|}
\hline
\multirow{2}{*}{\diagbox[width=16.5em]{\textbf{Training set}}{\textbf{Test set}}}
& \multicolumn{2}{c|}{\textbf{VieSpeaker}} \\
\cline{2-3}
& \textbf{E} & \textbf{H} \\
\hline\hline
VoxCeleb2 & 7.02 & \underline{12.95} \\
Vietnam-Celeb-T & 5.92 & 23.14 \\
VoxVietnam-T & 7.99 & 26.61 \\
VieSpeaker-T & \underline{2.40} & 13.45 \\
VoxCeleb2 ft. VieSpeaker-T & \textbf{1.81} & \textbf{9.83} \\
VoxCeleb2 ft. Vietnam-Celeb-T & 3.89 & 16.38 \\
VieSpeaker-T ft. Vietnam-Celeb-T & 3.24 & 19.29 \\
VoxCeleb2 ft. VoxVietnam-T & 5.50 & 22.14 \\
VieSpeaker-T ft. VoxVietnam-T & 3.66 & 18.85 \\
\hline
\end{tabular}
\end{table}
To support the experimental protocols, the dataset is divided into disjoint training and evaluation subsets. From the total 4,715 speakers, 715 speakers are randomly selected for evaluation using stratified sampling to preserve the original genre distribution. This evaluation pool is substantially larger than those of previous Vietnamese speaker recognition datasets, enabling more stable and statistically reliable performance estimation while reducing the risk of overfitting to a limited set of identities or trials. The remaining 4,000 speakers form the training subset, denoted as VieSpeaker-T.

Two evaluation protocols with different difficulty levels are constructed. VieSpeaker-E (Easy) forms positive pairs from utterances of the same speaker within a single video session, while negative pairs are sampled from different speakers. VieSpeaker-H (Hard) increases difficulty by forming positive pairs across different videos of the same speaker, introducing cross-session variability. In this setting, acoustically similar speaker pairs are preferentially selected to create more challenging negative trials. The final statistics of all subsets are summarized in Table~\ref{tab:subset}.

\section{Experiments}
\label{sec:experiments}
\subsection{Experimental setup}
All experiments are conducted with the WeSpeaker \cite{wespeaker} framework using the ECAPA-TDNN architecture with 1024-channel encoder blocks. During training, input audio is randomly cropped into 3-second segments. We extract 80-dimensional log Mel-filterbank features with a 25 ms frame length and 10 ms frame shift. The model is trained using Additive Angular Margin Softmax \cite{arcface} loss, with MUSAN \cite{musan} and RIRs \cite{rir} augmentation applied at a probability of 60\%. Training runs for 150 epochs on a single NVIDIA Tesla V100 GPU with a batch size of 128. Performance is reported using Equal Error Rate (EER), where lower values indicate better verification accuracy.

To analyze the impact of training data and pretraining strategies, we conduct two groups of experiments. First, models are trained from scratch on different datasets (VieSpeaker-T, Vietnam-Celeb-T, VoxVietnam-T, and VoxCeleb2) to compare their effectiveness as standalone training corpora. Second, we evaluate large-scale pretraining by using VoxCeleb2 and VieSpeaker as source datasets and finetuning them on Vietnamese benchmarks. Finetuning is performed for 40 epochs with a batch size of 64 on the respective target training sets.

\subsection{Experimental results}
Table~\ref{tab:cross_eval} reports EER (\%) on the Vietnam-Celeb and VoxVietnam benchmarks under different training and pretraining settings, providing a comprehensive comparison of models trained on existing Vietnamese datasets and the proposed VieSpeaker. 

When trained from scratch, the model trained on VieSpeaker-T achieves competitive results on Vietnam-Celeb-E/H while delivering the strongest performance on the VoxVietnam benchmark, outperforming models trained on the remaining datasets. The advantage of VieSpeaker is more evident in the pretraining setting: despite being roughly one-third the size of VoxCeleb2, it consistently yields stronger downstream performance. After finetuning on Vietnam-Celeb-T, the VieSpeaker-pretrained model achieves 5.45\% and 6.74\% EER on Vietnam-Celeb-E and H, corresponding to relative reductions of approximately 5.9\% and 2.5\% compared to VoxCeleb2 pretraining, and 16.5\% and 14.9\% compared to training from scratch. A similar trend is observed on the VoxVietnam benchmark, where pretraining on VieSpeaker yields slightly improved performance on VoxVietnam-E and comparable results on VoxVietnam-H compared to VoxCeleb2 pretraining.

Table~\ref{tab:vs_eval} reports results on the VieSpeaker benchmark. Although models trained on existing Vietnamese datasets achieve reasonable performance on VieSpeaker-E, they struggle on the harder VieSpeaker-H protocol, suggesting limited coverage of challenging scenarios in these corpora. In contrast, training on VieSpeaker-T from scratch substantially improves performance on both subsets, achieving relative EER reductions of 59.5\% and 41.9\% compared to the strongest prior Vietnamese dataset. Further gains are obtained when initializing from VoxCeleb2, reaching the best EER of 1.81\% and 9.83\% on VieSpeaker-E and H, respectively. We also compare the pretraining effectiveness of VoxCeleb2 and VieSpeaker for Vietnamese datasets on the VieSpeaker benchmark, where VieSpeaker pretraining consistently provides stronger improvements. Taken together with the results in Table~\ref{tab:cross_eval}, these findings highlight the effectiveness of VieSpeaker both as a training corpus and a pretraining resource for Vietnamese speaker recognition.

\section{Conclusion}
\label{sec:conclusion}
In this work, we introduced VieSpeaker, a large-scale Vietnamese speaker recognition dataset built using a face-independent pipeline. Our approach integrates speaker diarization and LLM-based identity reasoning to enable scalable identity annotation without relying on visual cues. VieSpeaker comprises 4,715 speakers and over 900 hours of speech from diverse real-world sources, representing a significant expansion in scale and diversity for Vietnamese speech resources. Experimental results suggest that models trained on VieSpeaker demonstrate promising improvements under challenging conditions. Furthermore, when used for pretraining, the dataset shows potential in enhancing cross-dataset generalization and downstream adaptation. These findings indicate that our face-independent pipeline opens a new avenue for large-scale data collection from public media, which can be combined with existing vision-based methods in the future to further broaden data sources. To foster ongoing research and facilitate the development of more robust models, the full dataset is made freely accessible to the community via Hugging Face\footnote{\url{https://huggingface.co/datasets/hustep-lab/VieSpeaker-Dataset}}. Ultimately, VieSpeaker serves as a solid foundation for advancing more robust Vietnamese speaker recognition systems across broader speech genres.

\section{Acknowledgments}
This research was funded by the Ministry of Education and Training of Vietnam under project code CT2025.EA.BKA.04.
\section{Generative AI Use Disclosure}
During the preparation of this manuscript, the authors utilized ChatGPT strictly for editing and polishing the text, ensuring that the AI tool was not used to produce any significant part of the manuscript. We thoroughly reviewed all suggestions and remain fully responsible and accountable for the final content of this work. Additionally, as detailed in Section~\ref{sec:pipeline}, a generative AI model (\texttt{gemini-2.5-pro}) was systematically utilized as a computational component within the proposed dataset construction pipeline for speaker identity reasoning.
\bibliographystyle{IEEEtran}
\bibliography{mybib}

\end{document}